# Sustainable low-cost method for production of High entropy alloys from alloy scraps

Karthikeyan Hariharan[#](0000-0003-3899-500X), K Sivaprasad[*](0000-0003-4358-3659)

Advanced Materials Processing Laboratory, Department of Metallurgical and Materials Engineering, National Institute of Technology, Tiruchirappalli, India 620015

[#]Present Address: Fontana Corrosion Centre, Department of Materials Science and Engineering, The Ohio State University, Columbus, OH, USA 43210

[*]Corresponding Author(ksp@nitt.edu); Telephone: +91-431-2503466

## Acknowledgements

The authors thank the Department of Materials Science and Engineering at The Ohio State University for providing access to the HEA database of the ThermoCalc software.

KH thanks Dr. Soumya Sridar, Hariharan Sriram, Vignesh Karunakaran and Prahnaov Arun for the useful technical discussions.

## Abstract

In this communication, we propose a sustainable way to produce high entropy alloys (HEAs) from alloy scraps called "Alloy mixing". We successfully demonstrate this method using a near-equimolar CrCuFeMnNi HEA. Alloy scraps (304L stainless steel (SS), Nichrome 80 and electrical wire grade Copper) obtained from various sources were melted together using vacuum arc melting along with minor additions of Mn and Cr to achieve the equiatomic composition. The alloy was characterized using X-ray Diffraction (XRD) and Scanning Electron Microscopy (SEM), which confirmed that the alloy produced through "Alloy mixing" exhibits a microstructure similar to that of the alloy with the same composition produced through conventional melting of pure elements. Property Calculation module on ThermoCalc was used to compare the yield strength of the conventional alloy and the alloy with impurities which indicated a 50% increase in yield strength. An uncertainty quantification analysis with 1000 alloy compositions with varying impurity contents indicates that the yield strength is strongly dependent on the impurity content. The cost analysis revealed that "Alloy mixing" would lead to a significant reduction in fabrication costs. These results are promising in the context of the commercialization of HEAs while offering a way to address the alloy scrap recycling problem.

**Graphical Abstract**

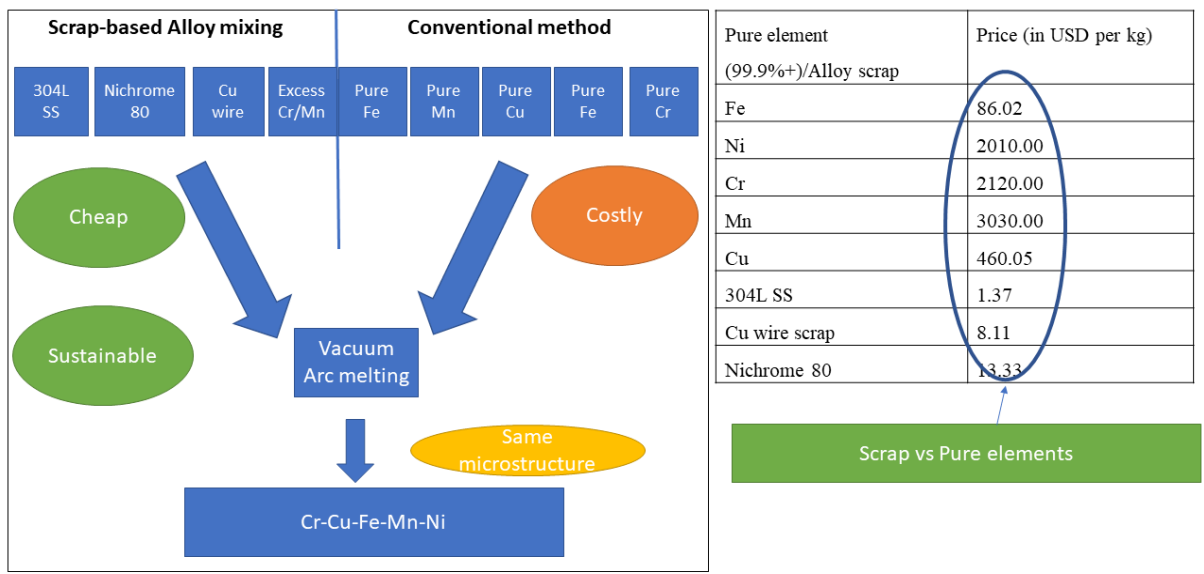

**1. Introduction**

The data published by the US Environmental Protection Agency [1] indicated that Municipal Solid Waste (MSW) in the US alone included 34.69 million tons of metallic scraps in the year 2018, of which only 34.9% was recycled. Recycling has been shown to reduce cost [2] and energy consumption [3] significantly as the primary metal production process is both cost and energy-intensive in nature. There have been several efforts in recent years to recycle metallic materials, especially in the steel industry. However, large quantities of metallic scraps remain as waste. Hence, we need more avenues for recycling metallic alloy scraps.

High entropy alloys (HEAs) are a new class of alloys with five or more elements in near equal proportions. [4][5] This new class of materials has attracted the alloying community due to its marked deviation from established norms of alloy design based on a single principal element. Over the years, researchers have found that HEAs can be tailored to possess several unique properties such as excellent fracture toughness [6], corrosion resistance [7] and catalytic properties [8]. Despite these developments on the alloy design front, the real-life application of these alloys is limited by their high cost as they are typically produced by vacuum melting of pure elements.

In this work, we propose a strategy named "Alloy mixing" to produce HEAs which can serve as a solution to the above mentioned problems. We melted alloy scraps together instead of melting pure metals to produce HEA, thereby reducing the cost of production significantly while enabling the recycling of metallic scraps. We demonstrate the feasibility of this technique using a CrCuFeMnNi HEA; the composition was chosen due to the easy availability of raw materials, and it has already been produced through the conventional process with its microstructure being well-characterized.

## 2. Materials and Methods

The precursor materials used in this study are 304L stainless steel (from "broken" tensile bars from a laboratory), Nichrome 80 (from "used" furnace coil), electrical grade Cu (from Cu wires). In addition to this, small quantities of 99.9% Mn and Cr were used to achieve the equiatomic composition. The alloys sourced from scraps were mechanically abraded on a belt grinder to remove any surface contamination.

An ingot weighing 30g was melted under a high vacuum in an Ar atmosphere on a vacuum arc furnace with a Tungsten electrode. The sample was re-melted atleast five times to ensure chemical homogeneity.

XRD analysis using a Cu-Kα source was performed for the as-cast sample for phase identification. The sample was metallographically prepared, and the microstructure was characterized using a SEM with a Field emission gun (FEG). In order to study the partitioning of elements between different phases, Energy dispersive spectroscopy (EDS) analysis was performed during the characterization under the SEM.

To evaluate the effect of impurities from the scrap on the yield strength of the alloy, the yield strength model included in the ThermoCalc (Version 2021b) Property calculation module. We compared the yield strength of the alloy without any impurities (conventionally fabricated alloy) with the yield strength of alloy with impurities at 25ºC. In this study, we considered Si and C as the main impurities present in the alloy fabricated by alloy mixing. The nominal composition of the alloy with impurities is given as the mean composition in Table 1. We also created 1000 variations in impurity concentration about the mean alloy composition using the uncertainty quantification function available on ThermoCalc. The input parameters used to create the variation is listed in Table 1. The main purpose of this analysis was to ascertain the effect of variability in scrap composition on the yield strength of the alloy.

Table 1 Input parameters used to generate different compositions of alloys with varying impurity contents.

| Alloying element | Mean composition (in wt%) | Δ Min/Max |
| --- | --- | --- |
| Fe | 19.08 | 0 |
| Cr | 18.24 | 0 |
| Ni | 20.59 | 0 |
| Mn | 19.28 | 0 |
| Cu | 23.20 | 0 |
| Si | 0.5 | 0.4 |
| C | 0.02 | 0.03 |

During this study, we considered face-centered cubic phase as the matrix phase, as it is the phase with the highest volume fraction as known from the data on alloy with the same composition fabricated using conventional route [9]. We used a constant grain size of 200 μm for both pure and impure variants as they are fabricated under the same solidification conditions. (i.e. vacuum arc melting)

We also performed a cost analysis to compare the prices of the pure elements and the scrap to demonstrate the economic impact of this new technique.

## 3. Results and Discussion

The XRD analysis shows that the as-cast alloy has three constituent phases (2 face-centered cubic phases and a body-centered cubic phase) (Figure 1). When we compare this pattern with the one reported by C. Li et al. [9] for the alloy with the same composition fabricated by the conventional method, the peaks for different phases matches well, clearly indicating that our method preserves the microstructure of the alloy. This is further confirmed by the SEM micrograph (Figure 2) of the alloy fabricated by alloy mixing, which captures the features such as flower-pot morphology of the 2$^{nd}$ phase and inter-phase boundary precipitation of the 3$^{rd}$ phase, which are well-documented for the alloy fabricated through the conventional route with the same composition.

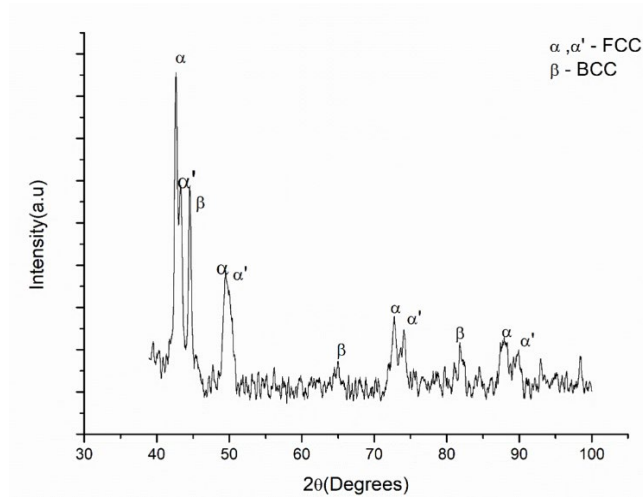

Figure 1 XRD pattern showing peaks corresponding to different phases present in the microstructure of the as-cast CrCuFeMnNi HEA fabricated using alloy mixing method.

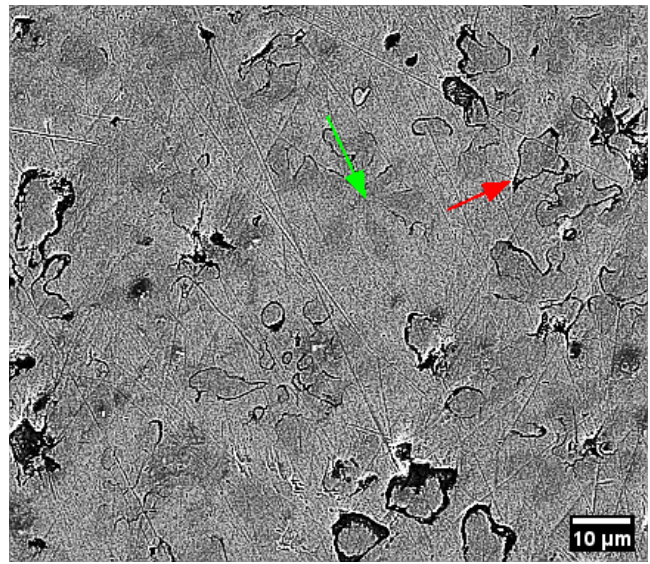

Figure 2 SEM secondary electron image showing the microstructure of the as-cast CrCuFeMnNi HEA fabricated through alloy mixing; the green arrow shows the β phase with flower-pot morphology, and the red arrow shows the α' phase on the phase boundary.

The EDS analysis (Figure 3) can be corroborated with the XRD data to ascertain the identity of each of the 3 phases present in the material. The Matrix phase-α (with a phase fraction of 57%) is an FCC phase with the presence of all elements except Cu. The β phase with the flower-pot morphology enriched with Cr is likely the BCC phase as Cr is known to stabilize the BCC structure over the FCC. The α' phase on the phase boundaries enriched with Cu is likely the second FCC phase as Cu has a tendency to segregate due to its positive enthalpy of mixing with the other alloying elements [10]

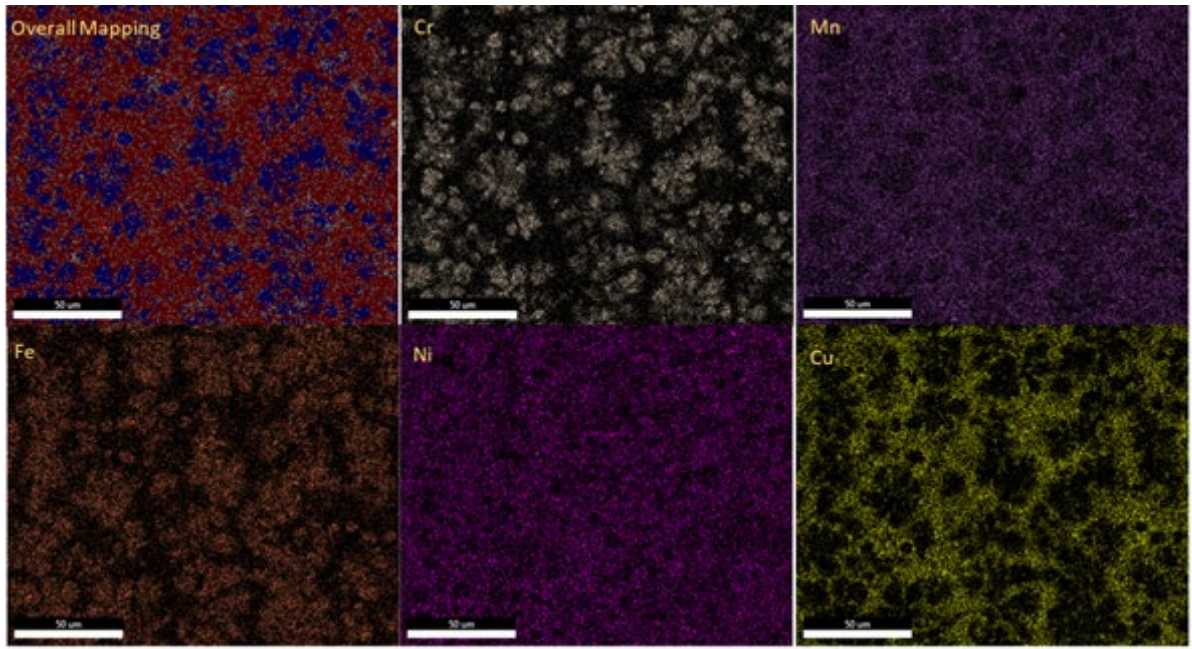

Figure 3 EDS maps showing different phases present and the distribution of different elements in the microstructure for the CrCuFeMnNi alloy produced using alloy mixing

The yield strength predicted by the ThermoCalc software for the alloy without impurities (the pure alloy) is found to be 135.55 MPa, while the alloy corresponding to the mean impurity concentration is 190.21 MPa which is 50% higher than the pure alloy. The difference can be fully attributed to the difference in solid solution strengthening between the two variants of the alloy, as we don't have any precipitation in both cases, and the grain size is assumed to be the same for both alloys. We should be careful in interpreting these results because we would have a difference in work hardening between the pure and impure variants, and we should also consider the effect of the impurity elements (i.e. Si and C) on the ductility of the alloys. However, we have results from the literature showing that Si improves work hardening rate by decreasing the Stacking fault energy (SFE)[11], and it also leads to uniform elongation and slightly enhanced ductility by the formation of refined dislocation cell-structure during plastic deformation[12][13]. The effect of carbon would be insignificant as it is present in very small amounts. Therefore, the impure alloys are expected to exhibit increased ultimate tensile strength along with good tensile ductility.

The results from the uncertainty analysis are presented below as a histogram (Figure 4a). It is interesting to note that the yield strength shows a large variation for small variation in the impurity content, which demonstrates the importance of controlling compositional variability in the scrap. It should be noted that some of the impure variants have a lower yield strength compared to that of the pure alloy, which shows that we should be careful in controlling the scrap composition. We plotted the yield strength as a function of Si content (Figure 4b), which shows that the yield strength varies linearly with Si content. This is interesting because Si, which is present as an impurity, helps in strengthening, indicating that impurities are not necessarily detrimental to the mechanical properties of the alloy. It is likely that Si improves the yield strength by solid solution strengthening. We didn't observe a clear trend with respect to the variation in Carbon content. It is due to the very low carbon content in the raw material. While the results from the strengthening model available in ThermoCalc are reliable, we need to conduct further experiments to get a complete understanding of the phenomena. However, these results are extremely useful in the context of guiding future experiments.

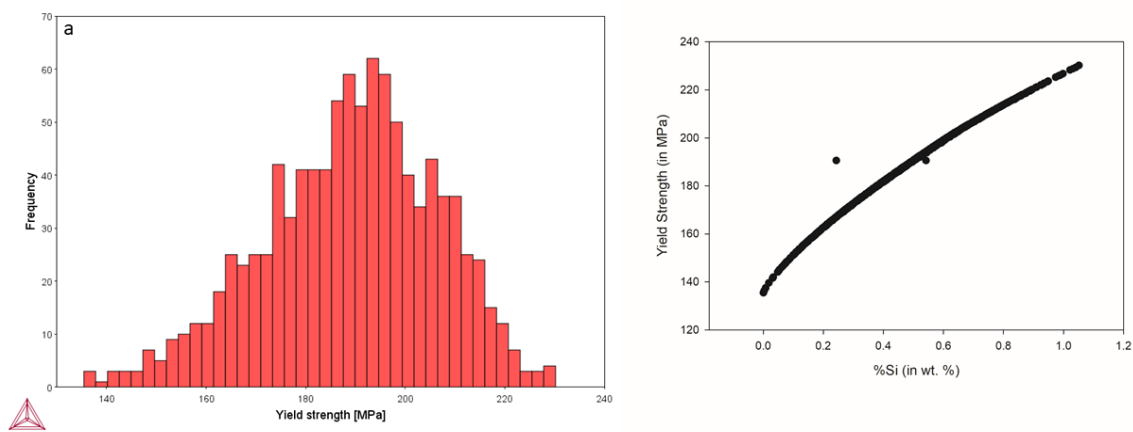

Figure 4 (a) Frequency distribution plot for the yield strength for alloys with varying impurity contents, (b) plot showing the variation of yield strength as a function of Si content

Table 2 lists the market price for the alloy scraps and the pure elements.[14][15] It is evident that alloy scraps are 100 times cheaper than pure elements. However, one must be aware of the fact that there is a cost associated with pre-treating the scraps. Hence, the actual price difference will be smaller than the apparent difference in the price listed. Nevertheless, the economic benefit that comes with "Alloy mixing" is significant in the context of HEA commercialization.

Table 2 List showing the price/kg of alloy scraps and pure element in USD

| Pure element (99.9%+)/Alloy scrap | Price (in USD per kg) |
|---|---|
| Fe | 86.02 |
| Ni | 2010.00 |
| Cr | 2120.00 |
| Mn | 3030.00 |
| Cu | 460.05 |
| 304L SS scrap | 1.37 |
| Cu wire scrap | 8.11 |
| Nichrome 80 scrap | 13.33 |

## 4. Conclusions

- Alloy mixing using scraps preserves the microstructure of the alloy as confirmed by characterization using XRD and SEM.
- The yield strength of the impure alloy was 50% higher than the conventionally manufactured alloy, indicating possible additional impact on the solid solution strengthening from the impurity elements, especially Si.
- The uncertainty quantification with respect to variation in impurity content indicated that the yield strength showed a large variation (nearly 100MPa) as a function of impurity concentration. It also revealed that if we don't control the scrap composition properly, we may end up degrading the properties.
- Our cost analysis revealed that alloy mixing could lead to a significant reduction in fabrication costs.

Hence, Alloy mixing serves as a promising sustainable and cost-effective method for the fabrication of HEAs, which would enable the commercialization of HEAs while serving as a feasible avenue for alloy scrap recycling. Future research must be focused on confirming the computational results with experiments and designing methods to effectively control the incoming scrap composition and establish robust scrap pretreatment cycles.

**Conflict of Interest:** The authors declare that they have no conflict of interest.

**References**

1. United States Environmental Protection Agency (2021) Advancing Sustainable Materials Management: 2018 Tables and Figures Assessing Trends in Material Generation and Management in the US;2021 ASI 9214-6.

2. Broadbent C (2016) Steel's recyclability: demonstrating the benefits of recycling steel to achieve a circular economy. Int J Life Cycle Assess 21:1658-1665. doi: 10.1007/s11367-016-1081-1.

3. Manabe T, Miyata M, Ohnuki K (2019) Introduction of Steelmaking Process with Resource Recycling. J Sustain Metall 5(3):319-330. doi: 10.1007/s40831-019-00221-1.

4. Yeh J-, Chen S-, Lin S-, Gan J-, Chin T-, Shun T-, Tsau C-, Chang S- (2004) Nanostructured High-Entropy Alloys with Multiple Principal Elements: Novel Alloy Design Concepts and Outcomes. Advanced engineering materials 6(5):299-303. doi: 10.1002/adem.200300567.

5. Cantor B, Chang ITH, Knight P, Vincent AJB (2004) Microstructural development in equiatomic multicomponent alloys. Materials Science and Engineering A: Structural Materials: properties,microstructure,processing 375-377:213-218. doi: 10.1016/j.msea.2003.10.257.

6. Li Z, Raabe D (2017) Strong and Ductile Non-equiatomic High-Entropy Alloys: Design, Processing, Microstructure, and Mechanical Properties. JOM 69:2099-2106. doi: 10.1007/s11837-017-2540-2.

7. Sahu S, Swanson OJ, Li T, Gerard AY, Scully JR, Frankel GS (2020) Localized Corrosion Behavior of Non-Equiatomic NiFeCrMnCo Multi-Principal Element Alloys. Electrochimica acta 354(C):136749. doi: 10.1016/j.electacta.2020.136749.

8. Tomboc GM, Kwon T, Joo J, Lee K (2020) High entropy alloy electrocatalysts: a critical assessment of fabrication and performance. Journal of Materials Chemistry A, Materials for Energy and Sustainability 8(3):14844-14862. doi: 10.1039/d0ta05176d

9. Li C, Li JC, Zhao M, Jiang Q (2009) Effect of alloying elements on microstructure and properties of multiprincipal elements high-entropy alloys. J Alloys Compounds 475:752-757. doi: 10.1016/j.jallcom.2008.07.124.

10. Troparevsky MC, Morris JR, Kent PRC et al (2015) Criteria for Predicting the Formation of Single-Phase High-Entropy Alloys. Physical review. X 5(1):011-041. doi: 10.1103/PhysRevX.5.01104.

11. Blinov VM, Bannykh IO, Lukin EI, Bannykh OA, Blinov EV, Chernogorova OP, Samoilova MA (2021) Effect of Substitutional Alloying Elements on the Stacking Fault Energy in Austenitic Steels. Russian metallurgy Metally 2021:1325. doi: 10.1134/S0036029521100086.

12. Xiong R, Liu Y, Si H, Peng H, Wang S, Sun B, Chen H, Kim HS, Wen Y (2020) Effects of Si on the Microstructure and Work Hardening Behavior of Fe–17Mn–1.1C–xSi High Manganese Steels. Metals and Materials International 3891-3904. doi: 10.1007/s12540-020-00846-y.

13. D. T. Llewellyn (1997) Work hardening effects in austenitic stainless steels. Materials Science and Technology 13:5, 389-400. doi: 10.1179/mst.1997.13.5.389.

14. Alfa-Aeser (2021) Price list for pure elements. https://www.alfa.com/en/pure-elements/. Accessed Oct 23, 2021.

15. iScrap (2021) Price list for metallic scarp. https://iscrapapp.com/prices/. Accessed Oct 16, 2021.